\documentclass[12pt,twoside]{article}
\usepackage[mathscr]{eucal}
\usepackage{amsmath,amsfonts,amssymb,amsthm,mathabx}
\usepackage{times}
\usepackage{pdfsync}
\usepackage{url}
\usepackage{hyperref}

\advance\textheight\topskip
\newcommand{\dd}{\partial}
\renewcommand{\d}{\partial}

\newcommand{\half}{\frac{1}{2}}

\newcommand{\ffrac}[2]{\raisebox{.5pt}%
  {\footnotesize$\displaystyle\frac{#1}{#2}$}\kern1pt}

\newcommand{\ddl}[2]{\ffrac{\dd #1}{\dd #2}}

\numberwithin{equation}{section} \makeatletter
\@addtoreset{equation}{section}

\DeclareFontFamily{OT1}{rsfs}{} \DeclareFontShape{OT1}{rsfs}{m}{n}{
<-7> rsfs5 <7-10> rsfs7 <10-> rsfs10}{}
\DeclareMathAlphabet{\mycal}{OT1}{rsfs}{m}{n}
\def\scri{{\mycal I}}%
\def\scrip{\scri^{+}}%

\newcommand*\xbar[1]{%
  \hbox{%
    \vbox{%
      \hrule height 0.5pt % The actual bar
      \kern0.3ex%         % Distance between bar and symbol
      \hbox{%
        \kern-0.0em%      % Shortening on the left side
        \ensuremath{#1}%
        \kern-0.0em%      % Shortening on the right side
      }%
    }%
  }%
}

\begin{document}

\title{Centrally extended BMS4 algebroid}

\author{Glenn Barnich}

\date{}

\def\mytitle{Centrally extended BMS4 Lie algebroid}

\pagestyle{myheadings} \markboth{\textsc{\small G.~Barnich}}{%
   \textsc{\small Centrally extended BMS4 Lie algebroid}}
 
\addtolength{\headsep}{4pt}

\begin{centering}

  \vspace{1cm}

  \textbf{\Large{\mytitle}}

  \vspace{1.5cm}

  {\large Glenn Barnich}

\vspace{1cm}

\begin{minipage}{.9\textwidth}\small \it  \begin{center}
    Physique Th\'eorique et Math\'ematique \\ Universit\'e Libre
    de Bruxelles and International Solvay Institutes \\ Campus Plaine
    C.P. 231, B-1050 Bruxelles, Belgium
 \end{center}
\end{minipage}

\end{centering}

\vspace{1cm}

\begin{center}
  \begin{minipage}{.9\textwidth}
    \textsc{Abstract}. We explicitly show how the field dependent
    2-cocycle that arises in the current algebra of 4 dimensional
    asymptotically flat spacetimes can be used as a central extension
    to turn the BMS4 Lie algebra, or more precisely, the BMS4 action
    Lie algebroid, into a genuine Lie algebroid with field dependent
    structure functions. Both a BRST formulation, where the extension
    appears as a ghost number 2 cocyle, and a formulation in terms of
    vertex operator algebras are introduced. The mapping of the
    celestial sphere to the cylinder then implies zero mode shifts of
    the asymptotic part of the shear and of the news tensor.
  \end{minipage}
\end{center}

\vfill

\thispagestyle{empty}
\newpage

\begin{small}
{\addtolength{\parskip}{-2pt}
 \tableofcontents}
\end{small}
\thispagestyle{empty}
\newpage

\section{Introduction}
\label{sec:introduction}

The holographic properties of three-dimensional anti de-Sitter gravity
are very special: the asymptotic symmetry algebra represents the two
dimensional conformal algebra with a definite prediction for the
classical central charge in the Dirac bracket algebra of the canonical
generators \cite{Brown:1986nw}, the solution space includes rotating
black holes \cite{Banados:1992wn}, and the dual theory on the
classical level is Liouville theory
\cite{Coussaert:1995zp}. Furthermore, using the classical central
extension in Cardy's formula allows one to reproduce the
Bekenstein-Hawking entropy of the BTZ black holes
\cite{Strominger:1998eq} (see also \cite{Carlip:1998qw}).

Whereas all these results have natural analogues in asymptotically
flat gravity at null infinity in three dimensions
\cite{Ashtekar1997,Barnich:2006avcorr,Barnich:2012aw,Barnich:2012rz,%
  Barnich:2012xq,Bagchi:2012xr,Barnich:2013yka}, the four-dimensional
case \cite{Bondi:1962px,Sachs1962a,Sachs1962} is more involved. One
reason is the non-integrability of the charges in the presence of
gravitational radiation \cite{Wald:1999wa}. Another source of
complication is due to the desire to include superrotations in the BMS
group \cite{nutku1992impulsive,Barnich:2016lyg,Strominger:2016wns} or
the associated algebra
\cite{Banks:2003vp,Barnich:2009se,Barnich:2010eb,Barnich:2011ct},
which goes together with allowing for suitable singularities in the
solution space on which superrotations act. Besides formal reasons
that drive one to include these additional symmetries, they have been
shown to lead to new physical applications such as the subleading soft
graviton theorem \cite{Cachazo:2014fwa,Kapec:2014opa} or the spin
memory effect \cite{Pasterski:2015tva,Nichols:2017rqr}. A naive
treatment of the charges associated to the extended algebra, that is
to say integrating regular solutions on the celestial sphere with
generators that have poles, generically leads to divergences
\cite{Barnich:2011mi,Flanagan:2015pxa}. One way to avoid dealing
directly with those is simply not to integrate over the celestial
sphere and to work with the local current algebra instead
\cite{Barnich:2013axa}. For some problems this is however not good
enough as they do require a well-defined integration, or more
precisely, a suitable moment map from solution space into the dual of
the symmetry algebra. One such problem concerns the central charge.

The aim of the present paper is to clarify formal aspects of the field
dependent central extension that appears in the modified Dirac bracket
algebra of charges and currents \cite{Barnich:2011mi,Barnich:2013axa},
by explicitly constructing the centrally extended Lie algebroid that
comes with the $\mathfrak{bms}_4$ algebra and its action on the free
data at null infinity.

More precisely, we first reformulate the field dependent central
extension as a local BRST $2$-cocycle, in much the same way the
Adler-Bardeen non-abelian gauge anomaly can be reformulated as the
BRST $1$-cocycle that appears in the transgression from the
characteristic class ${\rm Tr}\, F^3$ to the primitive element
${\rm Tr}\, C^5$ (see
e.g.~\cite{Stora:1983ct,zumino1985chiral,jackiw1985topological,Faddeev1984}).
Despite the BRST-type formulation, we do not imply that we are dealing
with gauge symmetries. Rather, as in the three-dimensional case where
the asymptotic symmetries of the gravitational/Chern-Simons theory
become the global symmetries of the dual Wess-Zumino-Witten or
Liouville-type theory, we consider BMS$_4$ as the global symmetry
group of a suitable dual theory. The BRST formulation here is just a
convenient way to encode Lie algebra or algebroid cohomology.

For the above considerations, the associated local functionals are
formal in the sense that they are given by equivalence classes of top
forms up to exact ones, which means that one disregards all boundary
terms that come from integrations by parts. When looking for concrete
realizations, one is led towards formulations in terms of vertex
operator algebras of conformal field theories where spatial integrals
correspond to taking residues (see e.g.~\cite{Kac1996} or
\cite{M.Schottenloher1945,guieu2007} for elementary introductions,
\cite{Kapec:2016jld} where related contour integrals have appeared in
the current context and \cite{Banerjee:2015kc,Banerjee:2016nio} for
related constructions applied to BMS$_3$).

Alternatively, as proposed in \cite{Barnich:2016lyg}, one may map
$\scrip$ to a cylinder times a line and explicitly realize the
centrally extended $\mathfrak{bms}_4$ algebroid using Fourier
analysis. The effect of mapping a gravitational solution from the
2-punctured Riemann sphere to the cylinder is a shift of the zero mode
of the asymptotic part of the shear, and thus also of the subleading
part of the angular metric, that is linear in retarded time. As a
consequence, this implies a constant shift of the zero mode of the
news, in direct analogy with the standard shift of the zero mode of
the energy momentum tensor in a conformal field theory.

Since it might not be widely known in the physics literature, we start
by briefly recalling the general framework for central extensions in a
Lie algebroid/Lie-Rinehart pair (see
e.g.~\cite{Mackenzie2005,LojaFernandes2006}) before applying the
construction to the case of interest.

\section{Local description of a Lie algebroid}
\label{sec:local-descr-lie}

Consider an algebra of functions $A$ in variables $\phi^i$ with
elements denoted by $f(\phi)$ and a vector space $\mathfrak g$
generated over $A$ by a set $e_\alpha$, with elements denoted by
$\xi=\xi^\alpha(\phi)e_\alpha$. The vector space $\mathfrak g$ is
turned into a Lie algebra by defining
\begin{equation}
  \label{eq:1}
[e_\alpha,e_\beta] =f^\gamma_{\alpha\beta}(\phi)e_\gamma,
\end{equation}
\begin{equation}
  \label{eq:4}
  [e_\alpha,f(\phi)]=R^i_\alpha(\phi)\partial_i f,
\end{equation}
where $\d_i=\frac{\partial}{\partial \phi^i}$,
by extending the bracket using skew-symmetry and the Leibniz rule, and by 
requiring that 
\begin{equation}
  \label{eq:2} 2R^i_{[\alpha}\partial_i
R^j_{\beta]}=f^\gamma_{\alpha\beta}R^j_\gamma,
  \end{equation}
 \begin{equation} R^i_{[\gamma}\d_i
f^\epsilon_{\alpha\beta]}=f^\epsilon_{\delta[\gamma}f^\delta_{\alpha\beta]},\label{eq:3}
\end{equation}
where square brackets denote skew-symmetrization of the included
indices of the same type, divided by the factorial of the number of
these indices. In other words, all Jacobi identities hold when using
the rules and conditions \eqref{eq:2} and \eqref{eq:3}. Alternatively,
instead of \eqref{eq:4}, one can define
$\delta_\xi f=\xi^\alpha R^i_\alpha\d_i f$ and 
\begin{equation}
[\xi_1,\xi_2]=(\xi_1^\alpha\xi^\beta_2f^\gamma_{\alpha\beta}
+\delta_{\xi_1}
\xi^\gamma_2-\delta_{\xi_2}{\xi^\gamma_1})e_\gamma\label{eq:24},
\end{equation}
with 
\begin{equation}
  \label{eq:49}
  [\delta_{\xi_1},\delta_{\xi_2}]=\delta_{[\xi_1,\xi_2]}. 
\end{equation}
Note that here and below, we will systematically use the notation
\begin{equation}
\delta_\xi f=[\xi,f]\label{eq:72}.
\end{equation} 

Introducing Grassmann odd variables $C^\alpha$ and
$\partial_\alpha=\frac{\partial}{\partial C^\alpha}$, the graded space
of polynomials in these variables taking values in $A$ is denoted by
$\Omega^*$. Its elements are denoted by
\begin{equation}
  \omega=\sum_{p=0}\frac{1}{p!}\omega_{\alpha_1\dots \alpha_p}(\phi)C^{\alpha_1}\dots
  C^{\alpha_p},\label{eq:22}
\end{equation}
where $\omega_{\alpha_1\dots \alpha_p}=\omega_{[\alpha_1\dots \alpha_p]}$. Equations
\eqref{eq:2} and \eqref{eq:3} are then equivalent to the requirement that
\begin{equation}
  \label{eq:5}
  \gamma=C^\alpha R^i_\alpha\d_i-\half C^\alpha C^\beta
  f^\gamma_{\alpha\beta}\partial_\gamma,
\end{equation}
is a differential on $\Omega^*$,
\begin{equation}
  \label{eq:6}
  \gamma^2=0. 
\end{equation}
The particular case where the $f^\gamma_{\alpha\beta}$ are constants
and do not depend explicitly on the fields is referred to as an action
algebroid. 

Note that in the case of interest below, fields and their derivatives
are relevant, so that these formulas have to be suitably interpreted
in, respectively extended to, the context of jet-bundles, see
e.g.~\cite{Andersonbook,Dickey:1991xa,Olver:1993,Barnich:2010xq}.

\section{Central extensions}
\label{sec:central-extension}

A trivial central extension is constructed by adding a
generator $Z$ to $\mathfrak g$, with $\widehat{\mathfrak g}$
consisting of elements
$\hat\xi=\xi^\alpha(\phi)e_\alpha+\xi^Z(\phi)Z$, while keeping
\eqref{eq:1} unchanged, while
\begin{equation}
  \label{eq:7}
  [Z,e_\alpha]=0=[Z,f(\phi)], 
\end{equation}
and the bracket again extended by skew-symetry and Leibniz rule.

Consider a 2-cocycle,
\begin{equation}
  \label{eq:8}
  \gamma \omega^2=0\iff
  R^i_{[\gamma}\d_i\omega_{\alpha\beta]}=\omega_{\delta[\gamma}f^\delta_{\alpha\beta]}.  
\end{equation}
This condition is equivalent to saying that $\widehat{\mathfrak g}$,
where \eqref{eq:1} is changed to
\begin{equation}
  \label{eq:9}
  [e_\alpha,e_\beta]
  =f^\gamma_{\alpha\beta}(\phi)e_\gamma+\omega_{\alpha\beta}(\phi)Z, 
\end{equation}
and all other relations are kept unchanged, is still a Lie algebroid.
In the case where $\omega^2$ is a coboundary,
\begin{equation}
\omega^2=\gamma\eta^1\iff
\omega_{\alpha\beta}=2R^i_{[\alpha}\d_i\eta_{\beta]}
-f^\gamma_{\alpha\beta}\eta_\gamma,\label{eq:10}
\end{equation}
this extended Lie algebroid is equivalent to the trivially extended
Lie algebroid by the change of generators
\begin{equation}
  \label{eq:11}
  e'_\alpha=e_\alpha-\eta_\alpha Z,\quad Z'=Z. 
\end{equation}
The differential of the extended Lie algebroid is 
\begin{equation}
  \label{eq:12}
  \hat\gamma=\gamma-\half C^\alpha
  C^\beta\omega_{\alpha\beta}\frac{\d}{\d C^Z}
\end{equation}
in the space of polynomials in $C^\alpha,C^Z$ with values in functions
of $\phi^i$. By construction, the $2$-cocycle $\omega^2$,
becomes trivial in the extended complex, $\omega^2=-\hat\gamma C^Z$.

For later use, we note that, if
$K_{\xi_1,\xi_2}=\omega_{\alpha\beta}\xi^\alpha_1\xi^\beta_2$,
$\eta_\xi=\eta_\alpha \xi^\alpha$, the cocycle condition \eqref{eq:8}
and the coboundary condition \eqref{eq:10} can
also be written as 
\begin{equation}
  \label{eq:17}
  K_{[\xi_1,\xi_2],\xi_3}-[{\xi_3},K_{\xi_1,\xi_2}]+{\rm
    cyclic}(1,2,3)=0, 
\end{equation}
\begin{equation}
  \label{eq:23}
K_{\xi_1,\xi_2}=[{\xi_1},\eta_{\xi_2}]-[{\xi_2},\eta_{\xi_1}]-\eta_{[\xi_1,\xi_2]},   
\end{equation}
and that the extension defined by \eqref{eq:9} is equivalent to
\begin{equation}
  \label{eq:16}
  [\hat \xi_1,\hat \xi_2]=[\xi_1,\xi_2]+K_{\xi_1,\xi_2}Z. 
\end{equation}

\section{BMS4 action algebroid}
\label{sec:bms4-acti-algebr}

We describe here relevant elements of the symmetry structure of
four-dimensional asymptotically flat spacetimes at null infinity. We
will adopt the point of view developed in
\cite{Barnich:2009se,Barnich:2010eb,Barnich:2011ct,Barnich:2011mi,%
  Barnich:2011ty,Barnich:2013axa,Barnich:2016lyg}, to which we refer
for more details and assume at the outset to be in the simplest case
for our purpose here, with future null infinity $\scrip$ taken as the
2-punctured Riemann sphere times a line.

The independent variables are $u,\zeta,\bar\zeta$. The variable $u$
is real, $\bar u=u$.  In this context, the $\mathfrak{bms}_4$ algebra
is parametrized by $T(\zeta,\bar\zeta)=\bar T$,
$Y(\zeta),\bar Y(\bar\zeta)$. It is the Lie algebra of vector fields
\begin{equation}
\xi=f\d_u+Y\d+\bar Y\bar\d,\label{eq:13}
\end{equation}
where $\d=\d_{\zeta}$, $\bar\d=\d_{\bar\zeta}$, 
\begin{equation}
f=T+\half u\psi,\quad 
\psi=\d Y+\bar\d\bar Y\label{eq:26}. 
\end{equation}
Writing
$[\xi_{T_1,Y_1,\bar Y_1},\xi_{T_2,Y_2,\bar Y _2}]=\xi_{\hat T,\hat
  Y,\hat{\bar Y}}$, this gives
\begin{equation}
  \label{eq:14}
 \widehat T=Y_1\d T_2-\half \d Y_1 T_2 -(1\leftrightarrow 2)+ {\rm
   c.c.}, 
\end{equation}
\begin{equation}
\widehat Y=Y_1\d Y_2-(1\leftrightarrow 2),\label{eq:19}
\end{equation}
where ${\rm c.c.}$ denotes complex conjugation and 
$\widehat{\bar Y}=\bar{\widehat Y}$.

The relevant fields are $\sigma(u,\zeta,\bar\zeta)$, its complex
conjugate $\bar \sigma$ and their derivatives. They correspond to the
asymptotic part of the complex shear, but for notational simplicity,
we have dropped the standard superscript $0$. On-shell, they encode
the subleading components of the angular part of the BMS metric. They
transform as
\begin{equation}
  \label{eq:15}
  -[{\xi},\sigma]=\big[f\d_u+Y\d+\bar Y\bar\d
  +\frac{3}{2}\bar\d\bar Y-\half \d Y\big]\sigma-\bar\d^2 f,
\end{equation}
with $\delta_{\xi}\bar\sigma=\xbar{\delta_{\xi}\sigma}$. Furthermore,
the transformation of the derivative of a field corresponds to the
derivative of the transformation of the field,
\begin{equation}
  \label{eq:25}
  [\xi, \d^k\sigma]=\d^k([\xi,\sigma]),\quad [\xi,\bar
  \d^k\sigma]=\bar \d^k([\xi,\sigma]),\quad
  [\xi,\d^k_u\sigma]=\d_u^k([\xi,\sigma]), 
\end{equation}
together with the complex conjugates of these relations. Note that
this implies in particular the following transformation law for the
news tensor $\dot\sigma=\d_u\sigma$, 
\begin{equation}
  \label{eq:34}
  -[\xi,\dot\sigma]=[f\d_u+Y\d+\bar Y\bar\d+2\bar\d
\bar Y]\dot\sigma-\half\bar\d^3\bar Y.
\end{equation}
It also follows that
$[{\xi_1},[{\xi_2},\sigma]]-[{\xi_2},[{\xi_1},\sigma]]=[[\xi_1,\xi_2],\sigma]$,
as required by \eqref{eq:2}. There are other fields on which
$\mathfrak{bms}_4$ acts non trivially, but they are passive in the
sense that they do not modify the commutators below. 

It follows from the computations in
\cite{Barnich:2011mi,Barnich:2011ty,Barnich:2013axa} that the
expression
\begin{equation}
  \label{eq:18}
  K_{\xi_1,\xi_2}= \int d\zeta \int
  d\bar\zeta\,\Big[
  \big(\sigma f_1\d^3Y_2-(1\leftrightarrow
  2)\big)+{\rm c.c.}\Big],  
\end{equation}
satisfies the cocycle condition \eqref{eq:17} provided that the
integral annihilates $\d$ and $\bar\d$ derivatives.

\section{BRST formulation}
\label{sec:brst-formulation}

Introducing the Grassmann odd fields $\eta(\zeta)$,
$\bar \eta(\bar\zeta)$, $C(\zeta,\bar\zeta)$ with $C$ real,
associated to $Y,\bar Y, T$, and the combination
$\chi=C+\frac{u}{2}(\d\eta+\bar\d\bar\eta)$, the BRST differential of
the BMS4 action algebroid is defined through
\begin{equation}
  \label{eq:27}
\begin{split}
  & \gamma \eta=-\eta\d \eta,\quad \gamma C=-\eta\d C+\half \d \eta C + {\rm c.c.},\\
  & \gamma \sigma=-\Big(\chi\d_u+\eta\d +\bar
  \eta\bar\d+\frac{3}{2}\bar\d\bar\eta-\half \d
  \eta\Big)\sigma+\bar\d^2 \chi.
\end{split}
\end{equation}
When changing variables and using $\chi(u,\zeta,\bar\zeta)$ instead of $C$, 
we have 
\begin{equation}
  \label{eq:29}
 \dot\chi=\half(\d\eta+\bar\d\bar\eta),\quad  \gamma \chi=-(\eta\d+\bar\eta
   \bar\d-\half\d\eta-\half\bar\d\bar\eta)\chi.
\end{equation}
The differential is
extended so as to commute with complex conjugation and all
derivatives. The expression corresponding to the integrand of
\eqref{eq:18} and the associated spatial components computed in
\cite{Barnich:2013axa} is given by
\begin{equation}
  \label{eq:18a}
\omega^{2,2}=d\zeta d\bar\zeta K^u -du d\bar \zeta K +du
  d\zeta \bar K, 
\end{equation}
where
\begin{equation}
  \label{eq:35}
  K^u=\chi(Q+\bar Q),\quad K =
  \eta(Q+\bar Q)+ \bar\d^3\bar\eta\d \chi,\quad Q=\d^3\eta\sigma.
\end{equation}
Introducing in addition 
\begin{equation}
N=\chi K,\quad \bar O=\eta\bar\eta (Q+\bar Q)+\eta\d^3\eta \bar\d
\chi-\bar\eta\bar\d^3\bar\eta \d \chi,
\end{equation}
and 
\begin{equation}
  \label{eq:36}
\begin{split}
\omega^{3,1}& =-(du\bar O-d\zeta \bar N+ d\bar\zeta N),\\
  \omega^{4,0}&=\eta\bar\eta \chi (Q+\bar Q)+\eta\d^3\eta \chi\bar \d
  \chi-\bar\eta\bar\d^3\bar\eta \chi\d \chi
\\
&=\eta\d^3\eta(\bar\eta
  \chi\sigma+ \chi\bar\d \chi)-\bar\eta\bar\d^3\bar\eta(\eta
  \chi\bar\sigma+ \chi\d \chi). 
\end{split}
\end{equation}
the relations
\begin{equation}
  \label{eq:32}
\begin{split}
  & \gamma
  Q=-\d(\eta Q)-\d_u (\chi Q)-\bar\d(\bar\eta Q+ \d^3\eta \bar \d
  \chi),\quad \gamma K^u=\d N+\bar \d \bar N, \\
  &  \gamma K=-\d_u N+\bar\d
  \bar O,\quad
 \gamma\bar O=-\d_u \omega^{4,0},\quad \gamma N=-\bar\d \omega^{4,0}, 
\end{split}
\end{equation}
allow one to easily derive the descent equations
\begin{equation}
\begin{split}
& \gamma \omega^{2,2}+d_H\omega^{3,1}=0\label{eq:44}, \\
& \gamma \omega^{3,1}+d_H\omega^{4,0}=0,\\
& \gamma \omega^{4,0}=0, 
\end{split}
\end{equation}
where $d_H=du\d_u+d\zeta\d +d\bar\zeta\bar\d$. It follows that
$\omega^{2,2}$ is a BRST cocycle modulo $d_H$ in ghost number $2$ and
form degree $2$. One way to show that this cocycle is non-trivial,
$\omega^{2,2}\neq \gamma \eta^{1,2}+d_H\eta^{2,1}$ is to show that
$\omega^{4,0}$ is non trivial, $\omega^{4,0}\neq \gamma
\eta^{3,0}$. This analysis will be completed elsewhere.

\section{Centrally extended BMS4 Lie algebroid}
\label{sec:centr-extend-bms4}

\subsection{Realization on the two-punctured Riemann sphere}
\label{sec:punct-riem-sphere}

Provided the integral annihilates spatial boundary terms, the
centrally extended Lie algebroid $\widehat{\mathfrak{bms}}_4$ is
defined by the commutators given in \eqref{eq:16}. More explicitly,
parametrizing $\widehat{\mathfrak{bms}}_4$ through
$(T(\zeta,\bar\zeta),Y(\zeta),\bar Y(\bar\zeta),V)$, with the
understanding that the elements in each slot can be multiplied by
functions of $\sigma,\bar\sigma$ and their derivatives, the
commutation relations \eqref{eq:14} and \eqref{eq:19} are completed by
\begin{equation}
  \label{eq:20}
  \hat V=K_{\xi_1,\xi_2}. 
\end{equation}

A concrete framework where multiplication is well defined and spatial
boundary terms can indeed be neglected is provided by vertex operator
algebras where one considers either polynomials with formal power
series or Laurent series and the integral is defined to select the
residue separately in $\zeta$ and $\bar\zeta$. For instance, a set-up
that accommodates singular solutions with delta function singularities
is to take for $Y,\bar Y,T$ Laurent polynomials, while
$\d_u^n\sigma$ are formal power series. 

In terms of the following generators for $Y,\bar Y,T$,
\begin{equation}
  \label{eq:59}
  l_m=-\zeta^{m+1}\d,\quad \bar l_m=-\bar\zeta^{m+1}\bar\d, \quad
  P_{k,l}=\zeta^{k+\frac{1}{2}}\bar\zeta^{l+\frac{1}{2}},
\end{equation}
the algebra reads
  \begin{equation}
\begin{split}
& [{l_m},{l_n}] = (m - n){l_{m + n}},\quad [{\bar l_m},{\bar l_n}] = (m
- n){\bar l_{m + n}},\\
  \label{eq:61}
& [{l_m},{P_{k,l}}] = (\half m- k){P_{m + k,l}},\quad [{\bar
  l_m},{P_{k,l}}] = (\half m - l){P_{k,m+l}},\\
& [{l_m},{\bar l_n}] = 0=[{P_{k,l}},{P_{o,p}}].
\end{split}
\end{equation}
It follows from \eqref{eq:15}, \eqref{eq:34} that the conformal
weights of $\sigma,\d_u\sigma$ are $(-\half,\frac{3}{2})$ and $(0,2)$
respectively, so that $u$ is of conformal weights
$(-\half,-\half)$. This leads to the expansions
\begin{equation}
\begin{split}
\d_u^n\sigma(u,\zeta,\bar\zeta)=\sum_{k,l}(\d_u^n\sigma)_{k,l}(u)
\zeta^{-k-\frac{n-1}{2}}\bar\zeta^{-l-\frac{n+3}{2}},\\ 
\d_u^n\bar
\sigma(u,\zeta,\bar\zeta)=\sum_{k,l}(\d_u^n\bar\sigma)_{k,l}(u)
\zeta^{-k-\frac{n+3}{2}}\bar\zeta^{-l-\frac{n-1}{2}}. \label{eq:21}
\end{split}
\end{equation}
Equation \eqref{eq:15}, \eqref{eq:34} and their
higher order time derivatives become
\begin{equation}
  \label{eq:65b}
\begin{split}
  &
  [l_m,(\d_u^n\sigma)_{k,l}]=(\frac{n-3}{2}m-k)(\d_u^n\sigma)_{m+k,l}
+\frac{m+1}{2}u(\d_u^{n+1}\sigma)_{m+k-\half,l-\half},\\ 
 & [\bar
 l_m,(\d_u^n\sigma)_{k,l}]=(\frac{n+1}{2}m-k)(\d_u^n\sigma)_{k,m+l}
+\frac{m+1}{2}u(\d_u^{n+1}\sigma)_{k-\half,m+l-\half}\\ & \hspace{4.5cm}-\half
m(m^2-1)(u\delta^0_n\delta^0_{k-\half}\delta^0_{m+l-\half}+\delta^1_n
\delta^0_{k}
\delta^0_{m+l}),\\
&[P_{k,l},(\d_u^n\sigma)_{o,p}]=-(\d_u^{n+1}\sigma)_{k+o,l+p}
+\delta^0_n(l^2-\frac{1}{4})\delta^0_{k+o}\delta^0_{l+p}.
\end{split}
\end{equation}
In these expansions, $m,n\in\mathbb{Z}$. One consistent choice, called
(NS) below,
which makes all the inhomogeneous terms above non-vanishing, is
$k,l,o,p\in\half +\mathbb{Z}$ when carried by $P$ or by an even number
of time derivatives of $\sigma$ and $k,l,o,p\in\mathbb{Z}$ when
carried by an odd number of time derivatives of $\sigma$. This means
that fields with odd conformal weights satisfy Neveu-Schwarz boundary
conditions and will be anti-periodic on the cylinder, while fields
with even conformal weights will be periodic on the cylider. Another
possibility (R) is to take $k,l,o,p\in \mathbb Z$ in all cases. Other
possibilities should of course be systematically explored.
 
Equation \eqref{eq:18} now implies the
following explicit expression for the central extension in terms of generators,
\begin{equation}
  \label{eq:45}
\begin{split}
  & K_{l_m,l_n}=\half u(m+1)(n+1)\sigma_{m+n-\half,-\half}[n(n-1)-m(m-1)],\\
  & K_{l_m,\bar l_n}=-\half u(m+1)(n+1)[\sigma_{m-\half,n-\half}m(m-1)-\bar\sigma_{m-\half,n-\half}n(n-1)],\\
  & K_{l_m,P_{k,l}}=\sigma_{m+k,l}m(m^2-1),\\
  & K_{\bar l_m,\bar l_n}=\half u(m+1)(n+1)\bar\sigma_{-\half,m+n-\half}[n(n-1)-m(m-1)],\\
  & K_{\bar l_m,P_{k,l}}=\bar\sigma_{k,m+l}m(m^2-1),\\
  & K_{P_{k,l},P_{o,p}}=0.
\end{split} 
\end{equation}
In case (NS), all these terms may be non-vanishing, while in case
(R) only $K_{l_m,P_{k,l}}$ and $K_{\bar l_m,P_{k,l}}$ may be.   

Even though it is not directly necessary for the construction of the
classical $\widehat{\mathfrak{bms}}_4$ Lie algebroid, we note in case
(NS) for instance, the
$\mathfrak{bms}_4$ Lie algebra itself may be encoded through the
series
\begin{equation}
  \label{eq:57}
\begin{split}
    & {J}(\zeta)=\sum_{m\in \mathbb Z}\zeta^{-m-2}l_m,\quad {\bar
      J}(\bar\zeta)=\sum_{m\in\mathbb Z}{\bar\zeta^{-m-2}}\bar l_m,\\
    & {P}(\zeta,\bar\zeta)=\sum_{k,l\in \half +\mathbb
      Z}\zeta^{-k-\frac{3}{2}}\bar\zeta^{-l-\frac{3}{2}}P_{k,l}, 
  \end{split}
\end{equation}
the commutation relations \eqref{eq:61} being equivalent to
\begin{equation}
\begin{split}
  \label{eq:58}
&  [{J}(\zeta),{J}(\omega)]
=(\delta(\zeta-\omega)D+2 D\delta(\zeta-\omega)){J}(\omega),\\
& [\bar J(\bar\zeta),\bar{J}(\bar \omega)]
=(  \delta(\bar\zeta-\bar\omega)\bar D+2 
\bar D\delta(\bar\zeta-\bar\omega))
\bar{J}(\bar\omega)
,\\
& [{J}(\zeta),{P}(\omega,\bar\omega)]=(\delta(\zeta-\omega)D
+\frac{3}{2}D\delta(\zeta-\omega))
{P}(\omega,\bar\omega)
, \\
& [\bar{J}(\bar\zeta), P(\omega,\bar\omega)]=
(\delta(\bar\zeta-\bar\omega)\bar D
+\frac{3}{2}\bar D\delta(\bar\zeta-\bar 
\omega)){
  P}(\omega,\bar\omega),\\
& [{J}(\zeta),\bar{J}(\bar\omega)]=0=[{J}(\zeta,\bar\zeta)
,{P}(\omega,\bar\omega)]. 
\end{split}
\end{equation}
with $D^k=\frac{1}{k!}\d^k_\omega$ and 
\begin{equation}
D^k\delta(\zeta-\omega)=\sum_{n\in\mathbb Z}\begin{pmatrix}n\\ k
\end{pmatrix}\zeta^{-n-1}\omega^{n-k}. 
\label{eq:62}
\end{equation}
As usual, in the space of formal distributions with values in the
universal enveloping algebra of $\mathfrak{bms}_4$, one can
write the singular parts as
\begin{equation}
  \label{eq:63}
\begin{split}
&  {J}(\zeta){J}(\omega)\sim \frac{D{J(\omega)}}{\zeta-\omega}
+\frac{2{J(\omega)}}{(\zeta-\omega)^2}, \\
& \bar {J}(\bar\zeta)\bar
    {J}(\bar\omega)\sim \frac{\bar D\bar{J}(\bar
      \omega)}{\bar\zeta-\bar\omega}
+\frac{2\bar{{J}}(\bar\omega)}{(\bar\zeta-\bar\omega)^2},\\
&{J}(\zeta){P}(\omega,\bar\omega)\sim\frac{D{P}(\omega,\bar\omega)}{\zeta-\omega}
+\frac{3{P}(\omega,\bar\omega)}{2(\zeta-\omega)^2},\\
&\bar{J}(\bar\zeta){P}(\omega,\bar\omega)\sim\frac{\bar D{P}(\omega,\bar\omega)}{\bar\zeta-\bar\omega}
+\frac{3{P}(\omega,\bar\omega)}{2(\bar\zeta-\bar\omega)^2},\\
& {J}(\zeta)\bar{J}(\xbar\omega)\sim 0\sim {P}(\zeta,\bar\zeta){P}(\omega,\bar\omega).
  \end{split}
\end{equation}
while 
\begin{equation}
  \label{eq:64}
  \begin{split}
    & [{J}(\zeta),\d_u^n{\sigma}(u,\omega,\bar\omega)]=(\delta(\zeta-\omega)D+
    D\delta(\zeta-\omega)[\frac{n-1}{2}+\frac{u}{2}\d_u])\d_u^n{\sigma}(u,\omega,\bar\omega), \\ 
& [\bar{J}(\bar\zeta),\d_u^n{\sigma}(u,\omega,\bar\omega)]=
(\delta(\bar\zeta-\bar\omega)\bar D+\bar
D\delta(\bar\zeta-\bar\omega)[\frac{n+3}{2}
+\frac{u}{2}\d_u])\d_u^n{\sigma}(u,\omega,\bar\omega)\\
& \hspace{8cm}-3(u \delta^0_n+\delta^1_n) \bar D^3\delta(\bar\zeta-\bar\omega)
,\\ 
&
[{P}(\zeta,\bar\zeta),\d_u^n{\sigma}(u,\omega,\bar\omega)]
=-\delta(\zeta-\omega)\delta(\bar\zeta-\bar\omega)\d_u^{n+1}\sigma(u,\omega,\bar\omega)\\
& \hspace{8cm} +2\delta^0_n\delta(\zeta-\omega)\bar D^2\delta(\bar\zeta-\bar\omega).
  \end{split}
\end{equation}
In the case of a suitable (free-field) representation of
$\mathfrak{bms}_4$ with locality conditions so that the various series
can be multiplied, one would write
\begin{equation}
  \label{eq:64a}
  \begin{split}
    & {J}(\zeta)\d_u^n{\sigma}(u,\omega,\bar\omega)\sim(\frac{1}{\zeta-\omega}D+
    \frac{1}{(\zeta-\omega)^2}[\frac{n-1}{2}+\frac{u}{2}\d_u])\d_u^n{\sigma}(u,\omega,\bar\omega), \\ 
& \bar{J}(\bar\zeta)\d_u^n{\sigma}(u,\omega,\bar\omega)\sim
(\frac{1}{\bar\zeta-\bar\omega}\bar D+\frac{1}{(\bar\zeta-\bar\omega)^2}[\frac{n+3}{2}
+\frac{u}{2}\d_u])\d_u^n{\sigma}(u,\omega,\bar\omega)\\
& \hspace{8cm}-3(u \delta^0_n+\delta^1_n) \frac{1}{(\bar\zeta-\bar\omega)^4}
,\\ 
&
{P}(\zeta,\bar\zeta)\d_u^n{\sigma}(u,\omega,\bar\omega)\sim-\frac{1}{\zeta-\omega}
\frac{1}{\bar\zeta-\bar\omega}\d_u^{n+1}\sigma(u,\omega,\bar\omega)\\
& \hspace{8cm} +2\delta^0_n\frac{1}{\zeta-\omega}\frac{1}{(\bar\zeta-\bar\omega)^3}.
  \end{split}
\end{equation}

\subsection{Realization on the cylinder}
\label{sec:realization-torus}

Alternatively, one may map $\scrip$ to a cylinder times a line and
consider Fourier series that can simply be multiplied under standard
assumptions.

As defined in \cite{Barnich:2010eb,Barnich:2011mi}, the
transformation laws of the $\mathfrak{bms}_4$ algebra under finite
superrotations are 
\begin{equation}
  \label{eq:47}
\begin{split}
&  Y'(\zeta')=Y(\zeta(\zeta'))\ddl{\zeta'}{\zeta},\quad \bar Y'(\bar \zeta')=\bar Y(\bar \zeta(\bar
  \zeta'))\ddl{\bar \zeta'}{\bar \zeta},\\ 
& T'(\zeta',\bar\zeta')=J^{-\half}
T(\zeta(\zeta'),\bar\zeta(\bar\zeta')),\quad
J=\ddl{\zeta}{\zeta'}\ddl{\bar\zeta}{\bar\zeta'},   
\end{split}
\end{equation}
while for the asymptotic part of the shear and its time derivatives,
equation (6.104) of \cite{Barnich:2016lyg} implies that  
\begin{equation}
  \label{eq:33}
  \d_{u'}^n\sigma'(u',\zeta',\bar\zeta')=(\ddl{\zeta}{\zeta'})^{\frac{n-1}{2}}
(\ddl{\bar\zeta}{\bar\zeta'})^{\frac{3+n}{2}}
[\d_u^n\sigma+\half(u\delta^0_n+\delta^1_n)\{\bar\zeta',\bar\zeta\}],\quad 
u'=J^{-\half}u.  
\end{equation}

The standard mapping from the cylinder to the 2-punctured Riemann
sphere is described by $\zeta=e^{\frac{2\pi}{L}\omega}$ with
$\omega=x_1+ix_2$, $x_2\sim x_2+ L$, coordinates on the cylinder.  Taking
$\zeta'=\omega,\bar\zeta'=\bar \omega$ in the above then gives
$l_m=-\frac{L}{2\pi}e^{\frac{2\pi}{L}m\omega}\partial_\omega$, $\bar
l_m=-\frac{L}{2\pi}e^{\frac{2\pi}{L}m\bar \omega}\partial_{\bar
  \omega}$, $u'=\frac{L }{2\pi}(\zeta\bar\zeta)^{-\half} u$,
$P_{kl}=\frac{L }{2\pi}e^{\frac{2\pi}{L} k\omega}e^{\frac{2\pi}{L}
  l\bar\omega}$, and the mode expansion
\begin{equation}
  \label{eq:53}
  \d_{u'}^n\sigma'(u',\omega,\bar \omega)
=(\frac{2\pi}{L})^{n+1}[(\d_u^n\sigma)_{k,l}(u)e^{-\frac{2\pi}{L}kw}e^{-\frac{2\pi}{L}l\bar
    w}]+(\frac{2\pi}{L})^2\frac{1}{4} (\delta^0_nu'+\delta^1_n).
\end{equation}

\section{Conclusion}
\label{sec:conclusion}

In this work, we have explicitly constructed a centrally extended Lie
algebroid associated to $\mathfrak{bms}_4$ on the two-punctured
Riemann sphere and the cylinder by suitably adapting the integration
rules and allowing for formal distributions.

Note that one could also have worked with appropriate distributions
directly on the celestial sphere (see
e.g.~\cite{Porter:1981cg,Galperin:1985bj,Galperin:1985va,Ivancovich1989,%
  Saidi1990,Saidi1992}). The point of view taken here consists in
first using transformation rules and invariance properties of various
quantities such as the Bondi mass aspect under conformal rescalings
\cite{Penrose:1984,Penrose:1986} to transpose everything to the
Riemann sphere before considering distributions.

Working out the details when starting from the celestial sphere
provides one with the normalizations for mass and angular momentum. In
this context, note that we have put the coefficient of the central
charge in \eqref{eq:18} to one. One should keep in mind however that
the correct normalization coming from the Einstein-Hilbert action is
$(16\pi G)^{-1}$ when integrated over the celestial sphere. For
instance, for asymptotically anti-de Sitter spacetimes in three
dimensions, it is this normalization that determines the precise
values $c^\pm={3l}/{2G}$ \cite{Brown:1986nw}. The correct
normalization is thus liable to play an important role in applications
such as Cardyology
\cite{Strominger:1998eq,Carlip:1998qw,Martinec:1998wm,Guica:2008mu,%
  Detournay:2012pc,Barnich:2012xq,Bagchi:2012xr} at null infinity
where $iu$ becomes a coordinate on the thermal circle, and so is the
shift in \eqref{eq:53} since the asymptotic part of the shear for the
Kerr black hole vanishes on the celestial sphere.

Apart from the concrete application considered in this work, the
current set-up paves the way for analyzing what happens to
gravitational solutions when replacing the celestial sphere by a
generic Riemann surface.

\section{Acknowledgements}
\label{sec:acknowledgements}

%\addcontentsline{toc}{section}{Acknowledgments}

This work is supported in part by the Fund for Scientific
Research-FNRS (Belgium), by IISN-Belgium, and by the Munich Institute
for Astro- and Particle Physics (MIAPP) of the DFG cluster of
excellence "Origin and Structure of the Universe". The author is most
grateful to C.~Troessaert for collaboration at an early stage and
thanks H.~Gonzalez, M.~Henneaux, B.~Oblak and S.~Lyakhovich for useful
discussions.

\addcontentsline{toc}{section}{References}

%\bibliography{C:/Users/Glenn/Dropbox/Literature/master}

\providecommand{\href}[2]{#2}\begingroup\raggedright\endgroup

\end{document}